\documentclass[graybox]{svmult}

\usepackage{type1cm}
\usepackage{makeidx}
\usepackage{graphicx}
\usepackage{multicol}
\usepackage[obeyspaces]{url}  
\usepackage[bottom]{footmisc}

\usepackage[numbers,sort]{natbib}

\usepackage{newtxtext}
\usepackage[varvw]{newtxmath}

\newcommand{\figref}[1]{Fig.~\ref{fig:#1}}
\newcommand{\tabref}[1]{Tab.~\ref{tab:#1}}

\newcommand{\ui}[1]{u_i^{(\text{#1})}}

\newcommand{\Nvox}{N_{\text{vox}}}
\newcommand{\Ncells}{N_{\text{cells}}}

\newcommand{\muprol}{\mu_{\text{prol}}}
\newcommand{\mudeath}{\mu_{\text{death}}}
\newcommand{\mudeg}{\mu_{\text{deg}}}

\newcommand{\muapopt}[1]{\mu_{\text{apopt,} #1}}

\newcommand{\MUapopt}{\mu_{\text{apopt}}}

\newcommand{\ecell}{e}

\newcommand{\Gprol}{g_{\text{prol}}}
\newcommand{\Eprol}{e_{\text{prol}}}

\newcommand{\gprol}[1]{g_{\text{prol,} #1}}
\newcommand{\eprol}[1]{e_{\text{prol,} #1}}

\newcommand{\cond}[1]{\text{cond}_{\text{#1}}}

\newcommand{\fitness}{R_{\text{pop}}}

\newcommand{\pext}{p^{(\text{ext})}}

\newcommand{\Omegacomp}{\Omega_{\text{comp}}}

\makeindex

\begin{document}

\title*{Modeling the hallmarks of avascular tumors}

\author{Erik Blom\orcidID{0009-0005-8141-6802} and \\
  Stefan Engblom\orcidID{0000-0002-3614-1732} and \\ 
  Gesina Menz\orcidID{0000-0002-3589-9824}}

\institute{Erik Blom, Stefan Engblom, and Gesina Menz \at Division of
  Scientific Computing, Department of Information Technology, Uppsala
  university, \email{erik.blom@it.uu.se}, \email{stefane@it.uu.se},
  \email{gesina.menz@it.uu.se}}

\maketitle

\abstract*{}

\abstract{We present a stochastic computational model of avascular
  tumors, emphasizing the detailed implementation of the first four
  so-called hallmarks of cancer: self-sufficiency in growth factors,
  resistance to growth inhibitors, avoidance of apoptosis, and
  unlimited growth potential. Our goal is to provide a foundational
  understanding of the first steps of cancer malignancy while
  addressing modeling uncertainties, thus bringing us closer to a
  first-principles grasp of this process. Preliminary numerical
  simulations illustrate the comprehensiveness of our perspective.}
  

\section{Introduction}
\label{sec:intro}

Cancer, a complex and multifaceted disease, has been characterized by
specific acquired capabilities underlying its progression
\cite{hanahan2000hallmarks}. These \emph{hallmarks of cancer},
introduced by Hanahan and Weinberg, aim to aid the understanding of
the disease complexity by providing a limited number of fundamental
characteristics which establish a \textit{conceptual scaffold} for
cancer research \cite{hanahan2022hallmarks}. The original six
hallmarks describe different ways in which cells evade normal
regulatory processes and attain features such as vascularization to
ensure the continued survival of a malignant tumor. They can occur in
a variable order and due to differing underlying cellular processes
depending on the specific cancer type.

By distilling cancer characteristics to a limited few overarching
processes, Hanahan and Weinberg’s hope was to enable researchers and
clinicians to specifically target aspects of cancer developments
through therapeutic measures. This goal of clinical application,
however, has only proven moderately successful due to the high
variability in pathways through which each individual hallmark can be
achieved \cite{hanahan2022hallmarks}.

Indeed, criticism against the hallmark concept has been voiced in this
context, stating that the low therapeutic success is due to focusing too
much on cell-based events and, thus, oversimplifying the
complexity of cancer as a tissue-level disease
\cite{sonnenschein2013critique}. Nevertheless, the hallmarks are a
widely accepted framework within the field of oncological research and
have given researchers a reasonably defined yet simple framework for
further research since their introduction.

One such avenue for further research is the use of modeling. There is
promising potential in furthering our understanding of a diverse range
of biological phenomena from carefully constructed computational
models, e.g., tumor growth laws \cite{benzekry2014classical} and
immune response \cite{eftimie2011interactions}. Such models supplement
biological experiments as they provide means to test hypotheses while
controlling for biological variability, can propose new experiments
and allow for new insights such as deciphering causal relationships
between different pathways \cite{brodland2015computational}.

Motivated by the utility and precision of mathematical models, we aim
to sharpen our understanding of cancer by interpreting its hallmarks
into an \textit{in silico} setting. Not only does this computational
version of the hallmarks serve as a way to sharpen their formulation
by making them quantitatively testable
\cite{brodland2015computational}, but we also assess the overall
usefulness and expressibility of our computational framework
\cite{engblom2018scalable}.

Our work in this paper revolves around implementing the first four
hallmarks of cancer. In \S\ref{sec:Methods} we describe our modeling
framework along with the specifics required to delineate each
hallmark. In \S\ref{sec:Results} we report some results from the
resulting numerical simulations and a concluding discussion is found
in \S\ref{sec:Discussion}.


\section{Methods}
\label{sec:Methods}

We interpret the first four hallmarks of cancer into an existing
stochastic and cell-based framework for avascular tumor growth
\cite{engblom2018scalable}, allowing for the study of the transition
from healthy homeostasis to the early stages of cancer. The
computational framework is summarized in \S\ref{subsec:DLCM} and the
hallmark-specific mechanisms are developed in
\S\ref{subsec:hallmarks_insilico}.

\subsection{Stochastic modeling framework}
\label{subsec:DLCM}

Our model is an extension of the avascular tumor model in
\cite{blom2023morphological}, which is based on the Darcy's Law Cell
Mechanics (DLCM) framework, in which cells are explicitly represented
and their states updated in a continuous-time Markov chain
\cite{engblom2018scalable}.

Let $\Omega$ denote the tissue domain populated with cells and let
$\Omegacomp$ denote the entire computational domain including the
space with no cells. In $\Omega$, cell migration for
cell density $u$ is governed by Darcy's law \cite{whitaker1986flow} for
fluid flow through porous media
\begin{equation}
   \boldsymbol{v} = - D\nabla p,
  \label{eq:cell_migration_law}
\end{equation}
where $p$ is the pressure and the constant $D$ is the the ratio of the
medium permeability to its dynamic viscosity. The pressure is governed
by a stationary heat equation with sources corresponding to crowding:
\begin{equation}
    \left. \begin{array}{rcl}
  -\Delta p &=& s(u), \\
  p &=& \pext - \sigma C, \qquad \text{ on } \partial\Omega
  \end{array} \right\}
  \label{eq:pressure_law}
\end{equation}
where $s(u)$ is a source function to be defined, $\pext$ is the
pressure outside the cell population, $C$ the population boundary
curvature, and $\sigma$ the surface tension coefficient. Cells are
represented as discrete units in a computational mesh, and pressure
sources arise only in voxels that exceed a \textit{carrying capacity},
here defined as one cell per voxel. While the original framework
considers cell proliferation and death as determined by an underlying
nutrient field, to limit the model complexity we here consider that
the cells always have sufficient nutrients for proliferation.

The domain is discretized into voxels $v_i$ for $i = 1, 2, ...\Nvox$
with total number of cells $\Ncells$. The number of cells in a voxel
is restricted to $u_i \in \{0, 1, 2\}$, and we define $s(u_i) = 1$ in
\eqref{eq:pressure_law} in $v_i$ if $u_i > 1$ and zero otherwise. A
cell can take on alternative phenotypes, e.g., necrotic or senescent,
and are then equipped with behavior distinct from regular cells, here
distinguished by a superscript as in $u_i^{(l)}$. Let $P(\omega)$
denote the rate of an event $\omega$, and let cell migration from
voxel $v_i$ to $v_j$ be denoted $i \rightarrow j$. The following
movement events based on \eqref{eq:cell_migration_law} are included
\begin{align}
  \label{eq:rates}
  \left. \begin{array}{rcl}
           P(i \rightarrow j; \; u_i \geq 1, \; u_j = 0) &=& D_1 q_{ij}\\
           P(i \rightarrow j; \; u_i > 1, \; u_j = 1) &=& D_2 q_{ij}
         \end{array} \right\}
\end{align}
where $D_1$ and $D_2$ are the Darcy coefficients for respective
movement type and $q_{ij}$ is the pressure gradient integrated over
the boundary shared between the voxels $v_i$ and $v_j$. All the rates
in this framework are interpreted as competing Poissonian events and
simulated by suitable algorithms.

\subsection{The hallmarks of cancer \textit{in silico}}
\label{subsec:hallmarks_insilico}

Although the framework in \S\ref{subsec:DLCM} supports basic tumor
models, it does not make the hallmarks of cancer explicit. To study
these we must therefore detail their mechanisms. A constraint that we
impose is that they should endow benign cell populations with the
capacity to maintain homeostasis.

We next describe how each of these four hallmarks are interpreted into
the model.

\subsubsection{Self-sufficiency in growth factors}
\label{paragraph:HMK1}

Healthy cells spend most of their time in a non-replicative state and
rely on growth signals to induce replication. Tumors, however, are
much less dependent on external signaling and are able to proliferate
in the absence of such environmental cues \cite{hanahan2000hallmarks}.

While the signaling network for initiating cell growth and
proliferation is complex \cite{hanahan2011hallmarks}, we shall consider only
short range growth signals induced within regions of low cell density such
that a population of healthy cells can return to homeostasis after a cell
loss. The growth signals are modeled as a quantity $g$ governed by
\begin{align}
  \left. \begin{array}{rcl}
           -\Delta g &=& s_g(v_i) - \gamma g \\
           g &=& 0, \quad \text{ on } \partial \Omegacomp
         \end{array} \right\}
         \label{eq:growth_signal_law}
\end{align}
with $s_g(v_i) = 1$ if $v_i$ is empty and zero otherwise. We include
the condition $g \geq \gprol{i}$ required for proliferation of cell
$i$, where $\Gprol$ and $\gamma$ together control the effective range
of the growth signal $g$. To estimate suitable values of $\Gprol$ and
$\gamma$ we solve \eqref{eq:growth_signal_law} in the radially
symmetric case when only a single center voxel is empty and for cell
diameter $h$. For large enough values of $\gamma$ and for $r \geq 0.5h$,
\begin{align}
   g(r) &= \frac{0.5h}{\sqrt{\gamma}} I_0(\sqrt{\gamma} h) K_0\left(
   \sqrt{\gamma} r \right),
\end{align}
in terms of modified Bessel functions of the first and second kind. To
arrive at an estimate, we impose that $\Gprol = g(h) = 4 g(2h)$, such
that the first layer of cells surrounding an empty voxel may
proliferate and that the second layer of cells requires roughly three
more empty voxels at equal distance to proliferate. For example, using
$h = 0.04$, this condition is satisfied when $\gamma \approx 700$ and
$\Gprol \approx 8 \times 10^{-5}$.

\subsubsection{Bypassing suppressive growth signals}

As described in \S\ref{paragraph:HMK1}, cells in healthy tissues are
assumed to spend the major parts of their lives in quiescence. This is
induced by the immediate surroundings of individual cells releasing
antigrowth signals to sustain tissue homeostasis. Tumor cells are able
to evade this external feedback by disrupting or completely blocking
pathways monitoring the received signals \cite{hanahan2000hallmarks}.

We introduce a local sensing mechanism by imposing the proliferation
condition $\ecell < \eprol{i}$ for cell $i$, where $\ecell$ is
defined as the weighted voxel edge ratio
\begin{align}
  \ecell = \frac{\sum_{j\in N_i} u_j E_{ij}}{\sum_{j\in N_i} E_{ij}},
  \label{eq:edge_ratio}
\end{align}
where $E_{ij}$ is the edge length shared between $v_i$ and $v_j$, and
$N_i$ is the set of all neighboring voxels to $v_i$. Thus, cell
proliferation is suppressed so long as a certain fraction of the
cell's boundary is adjacent to other cells.

\subsubsection{Avoiding apoptosis}

Apart from quiescence, healthy tissues also require apoptosis to
maintain tissue homeostasis. If cells experience high levels of stress
such as DNA damage or overcrowding, the apoptotic pathway can be
triggered to protect the tissue by reducing overall stress levels. It
is therefore advantageous for cancers to be able to evade apoptosis to
ensure continued proliferation even when cellular stress is high
\cite{hanahan2011hallmarks}.

We assume that all living cells are equally likely to sustain
spontaneous cell damage enough to initiate apoptosis: this is covered
by a small background death rate $\epsilon \muapopt{i}$ for each cell
$i$. We judge that signal imbalance occurs for cells $i$ with
$g < \gprol{i}$ or $\ecell \geq \eprol{i}$ in voxels where $u_i >
1$. In other words, a recently proliferated cell with incoming signals
which are in conflict with a proliferation event would trigger this
signal imbalance. Finally, we measure DNA damage by the deviation of
the first two cell-specific hallmark parameters $\theta_i = [\gprol{i},
\eprol{i}]$ from the their corresponding healthy values $\theta_0 = [\gprol{0},
\eprol{0}]$, and we assume that both signal imbalance and DNA damage induce
apoptosis at a rate proportional to $\muapopt{i}$ as specified in
\S\ref{paragraph:HMK4} below.

\subsubsection{Unlimited growth potential}
\label{paragraph:HMK4}

While the three previous hallmarks help tumor grow by decoupling cell
proliferation from external signals, an additional requirement for the
development of macroscopic tumors is the loss of replicative
limitations. Healthy cells generally have the capacity to proliferate
only a finite number of times. Cancer cells, however, are often able
to replicate infinitely often \cite{hanahan2000hallmarks,
  hanahan2011hallmarks}.

To achieve this in our model, we equip each cell $i$ with a number
$\eta_i$ such that cells with $\eta_i = 0$ can no longer
proliferate. After proliferation, each daughter cell inherits the
value of $\eta_i$ reduced by one, and we include a possibility of
mutating this value and the other hallmark phenotypes as described in
\S\ref{sec:Results} below.

Including proliferation and degradation events, all event rates thus read as
\begin{align}
  \label{eq:rates_HMK}
  & \hphantom{1cm} \left. \begin{array}{rcl}
           P(u_i \rightarrow u_i + 1; \; \cond{prol}) &=&
           \muprol u_i\\
           P(u_i \rightarrow \ui{necr}) &=& \muapopt{i} u_i \left(\epsilon +
           d(\theta_i, \theta_0)\right)\\
           P(u_i \rightarrow \ui{necr}; \; \cond{apopt} ) &=& \muapopt{i}
                                                              u_i, \\
           P(\ui{necr} \rightarrow \ui{necr} - 1) &=& \mudeg \ui{necr}, 
         \end{array} \right\} \\
  &\mbox{where } \cond{prol} \; \text{is} \; u_i < 2, \; g
  \geq \gprol{i}, \; \ecell < \eprol{i}, \; \eta_i > 0,\\
  &\mbox{and where } \cond{apopt} \; \text{is} \; u_i > 1, \; g < \gprol{i} \text{ or }
  \ecell \geq \eprol{i},
\end{align}
and $d(\cdot, \cdot)$ is the $l_2$-distance in the parameter space.
To simplify the implementation we impose $\ui{necr} \leq 1$ by letting
death events in voxels with $u_i>1$ simply remove the cell instead of
it becoming necrotic.

In this way the first four \textit{in silico} hallmarks are achieved
in the limit when $\Gprol \to 0$, $\Eprol \to 1$, $\MUapopt \to 0$,
and when $\eta$ is sufficiently large for the considered tumor
timescale. The model parameters are summarized in
\tabref{model_parameters}.

\begin{table}
  \centering
  \begin{tabular}{p{3cm}p{8cm}}
    \hline
    Parameter & Description \\
    \hline
    $D_1 = 1$ & Ratio medium permeability to dynamic viscosity
          $[f^{-1}l^{3}t^{-1}]$ \\
    $D_2 = 25$  & As $D_1$, but movement from doubly to singly occupied voxel   \\
    $\muprol = 1$ & Rate of cell proliferation $[t^{-1}]$ \\
    $\mudeath = 1$ & Rate of cell death $[t^{-1}]$ \\
    $\mudeg = 0.1$ & Rate of dead cell degradation $[t^{-1}]$\\
    $\sigma = 10^{-4}$ & Surface tension coefficient $[f]$ \\
    $\gamma = 500$ & Ratio of growth signal decay to diffusion rate $[l^{-2}]$ \\
    $\epsilon = 5 \times 10^{-3}$ & Background apoptosis rate factor $[-]$ \\
    $\gprol{0} = 10^{-4}$ & Healthy growth signalling threshold $[l^{-2}]$ \\
    $\eprol{0}$ = 1 & Healthy anti-growth signalling threshold $[-]$ \\
    $\muapopt{0}$ = 0.25 & Healthy apoptosis rate $[t^{-1}]$ \\
    $\eta_0$ = 20 & Healthy cell proliferation number $[-]$\\
    \hline
  \end{tabular}
  \caption{Cell population model parameters. The units $(t,l,f)$ correspond
  to units of time, length, and force, respectively.}
  \label{tab:model_parameters}
\end{table}


\section{Numerical results}
\label{sec:Results}

We focus here on a numerical illustration that provides insight into
the model's capacity to capture the transition from normal tissue to
malignancy, thus offering a fairly detailed depiction of the early
stages of cancer development.

We monitor the overall \emph{tumor reproduction number} $\fitness$
defined as the population mean of the cells' expected number $R_i$ of
proliferation events before death. To find $R_i$ for cell $i$, we
consider the equivalent branching process of a sequence of Bernoulli
trials with probability $q_i$ for a proliferation event and
$p_i \equiv 1-q_i$ for a death event that ends the process. We
identify the proliferation probability with the total instantaneous
proliferation- and death rates, $b_i$ and $d_i$, respectively,
according to $q_i = b_i/(b_i+d_i)$. However, we let $q_i = 0$ for the
final step of the sequence after $\eta_i$ proliferation events have
occurred to account for the known limit of each cell's growth
potential. The expected number of proliferation events of this
branching process is
\begin{align}
  \label{eq:fitness_i}
  R_i &= \sum_{j=0}^{\eta_i-1}jq_i^j(1-q_i) + \eta_i q_i^{\eta_i} =
        q_i\frac{1 - q_i^{\eta_i}}{1-q_i}. \\
  \intertext{Taking the population mean of this we arrive at the
  reproduction number}
  \label{eq:fitness}
  \fitness &= \Ncells^{-1}\sum_i \frac{b_i}{d_i} (1 - q_i^{\eta_i}).
\end{align}
We use the parameter values in \tabref{model_parameters} where we set
the initial hallmark parameters for all cells to respective
\textit{healthy} parameter values.  We include the probability
$\alpha_4 = 10^{-3}$ that $\eta_i$ increases by $1000$ at cell
division, and let the other three hallmark phenotypes mutate at cell
division to study the selection pressure on the hallmark
capabilities. The mutation process for the first three hallmark
phenotypes is a drift-free geometric Brownian motion in phenotype
space with diffusion coefficient equal to $\alpha_i = 0.2$ for
$i = 1, 2, 3$, rejecting values outside the domain of validity for
each phenotype. We induce a higher cellular turnover by imposing
necrosis at rate $\mudeath$ for cells in the central region
$r \leq 0.2$ of the population to increase the effective mutation
rates. The initial state is $u_i = 1$ everywhere in a circular domain
of unit radius, discretized by a Cartesian mesh with voxel size
$h = 0.04$.

\begin{figure}[h]
  \sidecaption
  \includegraphics[scale=1]{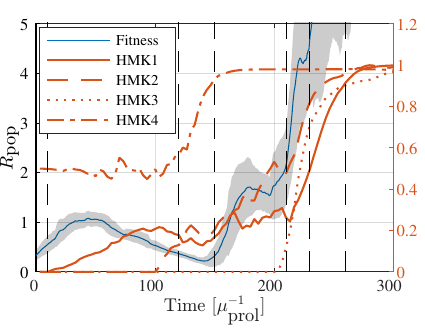}
  \caption{Population reproduction number (``fitness'') with the
    shaded grey area indicating $\pm 1$ standard deviation, together
    with the status of each hallmark during one simulation. Mutation
    rates for the second and third hallmark capabilities are activated
    after $t = 100$ and $200$, respectively. The hallmarks are
    represented by the population mean of the corresponding hallmark
    parameter mapped to $[0,1]$.}
\label{fig:hmks1}
\end{figure}

\figref{hmks1} summarizes the results of one simulation, including the
population means of respective hallmark parameter (denoted by bars)
mapped to $[0,1]$ via $1-\bar{g}_{\text{prol}}/\gprol{0}$,
$\bar{e}_{\text{prol}}/\eprol{0} - 1$,
$1-\bar{\mu}_{\text{apopt}}/\muapopt{0}$, and
$\bar{\eta}/(\bar{\eta} + \eta_0)$. The means are computed for the
population of cells within $r \leq 0.6$ to reduce the impact from
boundary effects. We observe a selection pressure towards the
activation of the hallmarks once the corresponding mutation is
allowed, but also that the DNA damage and signaling imbalance
mechanisms counteract hallmark progression. Hence the third hallmark
constitutes a significant threshold for malignancy in the model.

\begin{figure}[h]
    \sidecaption
  \includegraphics[scale=1]{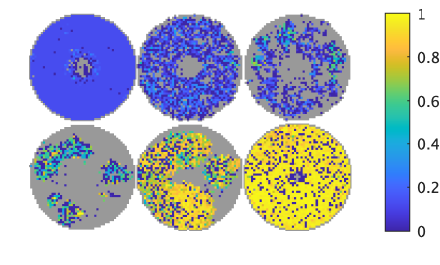}
  \caption{Mean hallmark status per cell at times corresponding to the
    vertical dashed lines in \figref{hmks1} at $t=10$, $120$, $150$,
    $210$, $230$, and $260$, respectively, from left to right, top to
    bottom. Grey indicates empty voxels.}
\label{fig:hmks1_spatial}
\end{figure}

\figref{hmks1_spatial} visualizes the spatial concentration of
hallmark phenotypes at selected times from the simulation. Thus, we
can localize where hallmark progression occurs to study the spatial
components (if any) of both the reproductive potential and the
selective pressure on the acquired phenotypes. For example, we observe
that while some regions in the population initially progress faster
towards hallmark achievements, progress ultimately spreads across the
entire population. We also observe that population decline and growth
correspond well with the proposed population reproduction number
$\fitness$.


\section{Discussion}
\label{sec:Discussion}

Our aim with the paper was two-fold: to provide a foundation for
understanding tumor evolution and to address ontological modeling
uncertainties, thereby progressing towards a more comprehensive
first-principles description of this intricate process.

To achieve our aim, we have interpreted the first four hallmarks of
cancer as mechanisms in a general framework for cell population
models. Through numerical experiments with locally exaggerated
mutation rates and cellular turnover we observe the selective pressure
within the studied time frame and a non-trivial progression towards
hallmark capabilities, both in terms of population mean and phenotype
spatial distributions. The results highlight the capacity of the model
to test the impact of precisely defined hallmark capabilities on the
cell population homeostasis, reproductive potential, and malignancy.

The results herein not only highlight the rich detail offered by
computational spatial models of tumorigenesis, but is also suggestive
of their potential to serve as pedagogic tools for approaching an
intuitive understanding of the dynamics of tumors. Our work emphasizes
the importance of balancing modeling granularity with a holistic view
to shed light on the fundamental steps towards malignancy.

\subsection{Availability and reproducibility}
\label{subsec:reproducibility}

The computational results and figures can be reproduced with release 1.4
of the URDME open-source simulation framework \cite{URDMEpaper}, available
for download at \url{www.urdme.org} (see the avascular tumor example and
the associated README in the DLCM workflow).



\ethics{Competing Interests}{This study was partially funded by the
  Swedish Research Council [grant number 2019-03471]. The authors have
  no conflicts of interest to declare that are relevant to the content
  of this chapter.}


\bibliographystyle{abbrvnat}
\bibliography{ENUMATH23}

\end{document}